\RequirePackage{amsmath}
\documentclass[runningheads]{llncs}

\usepackage[T1]{fontenc}
\usepackage[misc,geometry]{ifsym}
\usepackage{graphicx,amsfonts}

\begin{document}
%
%\title{Feature extraction on XAI maps to improve a model's performances: a study in MS lesion segmentation}
\title{Exploiting XAI maps to improve MS lesion segmentation and detection in MRI}
\titlerunning{XAI maps to improve model's performance}
\author{Federico Spagnolo\inst{1,2,3,4}\orcidID{0000-0001-9606-0400} \and
Nataliia Molchanova\inst{4,5}\orcidID{0000-0002-7211-8863} \and
Mario Ocampo Pineda\inst{1,2,3}\orcidID{0000-0003-2239-7355} \and
Lester Melie-Garcia\inst{1,2,3}\orcidID{0000-0001-5602-8916} \and
Meritxell Bach Cuadra\inst{5,6}\orcidID{0000-0003-2730-4285} \and
Cristina Granziera\inst{1,2,3}\orcidID{0000-0002-4917-8761} \and
Vincent Andrearczyk\inst{4}\orcidID{0000-0003-0793-5821} \and
Adrien Depeursinge\inst{4,7}\orcidID{0000-0002-2362-0304}\Letter}
\authorrunning{F. Spagnolo et al.}
\institute{Translational Imaging in Neurology (ThINk) Basel, Department of Medicine and Biomedical Engineering, University Hospital Basel and University of Basel, Basel, Switzerland \and
Department of Neurology, University Hospital Basel, Basel, Switzerland \and Research Center for Clinical Neuroimmunology and Neuroscience Basel (RC2NB), University Hospital Basel and University of Basel, Basel, Switzerland \and MedGIFT, Institute of Informatics, School of Management, HES-SO Valais-Wallis University of Applied Sciences and Arts Western Switzerland, Sierre, Switzerland \and CIBM Center for Biomedical Imaging, Lausanne, Switzerland \and Radiology Department, Lausanne University Hospital (CHUV) and University of Lausanne, Lausanne, Switzerland \and Nuclear Medicine and Molecular Imaging Department, Lausanne University Hospital (CHUV), Lausanne, Switzerland \\
\Letter \email{adrien.depeursinge@hevs.ch}}
\maketitle              % typeset the header of the contribution
\begin{abstract}

To date, several methods have been developed to explain deep learning algorithms for classification tasks. Recently, an adaptation of two of such methods has been proposed to generate instance-level explainable maps in a semantic segmentation scenario, such as multiple sclerosis (MS) lesion segmentation. In the mentioned work, a 3D U-Net was trained and tested for MS lesion segmentation, yielding an F1 score of 0.7006, and a positive predictive value (PPV) of 0.6265. The distribution of values in explainable maps exposed some differences between maps of true and false positive (TP/FP) examples. 
Inspired by those results, we explore in this paper the use of characteristics of lesion-specific saliency maps to refine segmentation and detection scores. We generate around 21000 maps from as many TP/FP lesions in a batch of 72 patients (training set) and 4868 from the 37 patients in the test set. 93 radiomic features extracted from the first set of maps were used to train a logistic regression model and classify TP versus FP.
On the test set, F1 score and PPV were improved by a large margin when compared to the initial model, reaching 0.7450 and 0.7817, with 95\% confidence intervals of [0.7358, 0.7547] and [0.7679, 0.7962], respectively.
These results suggest that saliency maps can be used to refine prediction scores, boosting a model's performances. 

%{The abstract should briefly summarize the contents of the paper in 150--250 words.}

\end{abstract}

\section{Introduction}

Multiple sclerosis (MS) is a demyelinating and autoimmune disease of the central nervous system, which increasingly affects the quality of life of relatively young people~\cite{koltuniuk2023}. A crucial magnetic resonance imaging (MRI) biomarker in diagnosing and monitoring MS is the presence of plaques (or lesions) in the white matter (WM), which are visible on fluid attenuated inversion recovery (FLAIR) and T1-weighted contrasts, such as the magnetisation-prepared rapid gradient echo (MPRAGE)~\cite{yang2022,thompson2017,hemond2018}.

Manual or semi-automatic annotation of such lesions is a tedious process, which has been automated in many tools based on deep learning (DL)~\cite{ma2022,commowick2023}. 
The ``black box'' nature of standard DL models~\cite{baselli2020} and the lack of clinical validation~\cite{spagnolo2023} have jeopardized the clinical integration of these tools. In this sense, research in explainable AI (XAI) could result as decisive to better understand and optimize the architecture and the performances of DL models \cite{kobayashi2024}. An exhaustive review of XAI models and applications published before 2023 can be found in~\cite{saranya2023}.

\begin{figure}[!ht]
\includegraphics[width=\textwidth]{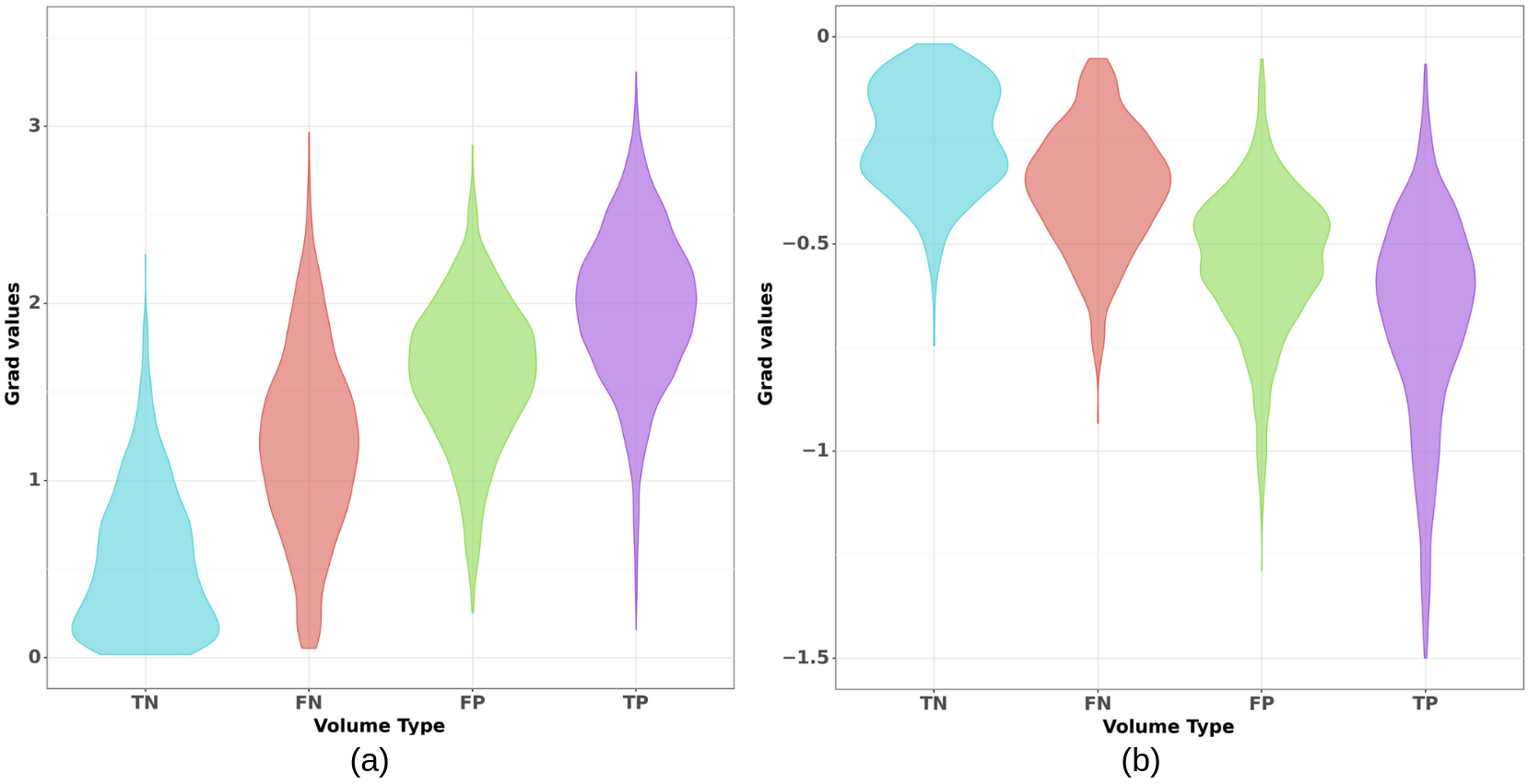}
\caption{Violin plots representing the distribution of saliency maps maximum (a) and minimum (b) values. The four distributions refer to true negative (TN), false negative (FN), false positive (FP) and true positive (TP) volumes. Figure retrieved from~\cite{spagnolo2024}.} \label{violinplots}
\end{figure}

However, XAI methods were not designed for segmentation tasks and, to date, no ad-hoc methods were capable to do so~\cite{arun2021,mahapatra2022}. To this end, two methods were recently developed in~\cite{spagnolo2024}, which adapt SmoothGrad~\cite{smilkov2017} and GradCAM++~\cite{chattopadhay2018} to provide instance-level explanation maps. Therefore, these two methods can generate separate (i.e., lesion-specific) explainable maps for distinct instances of a class, e.g., MS plaques. This is important to understand which parts of the image were responsible for the segmentation of a specific targeted lesion.
The distribution of maximum and minimum values in explainable (saliency) maps generated with the adapted SmoothGrad for true positive (TP) predictions was compared to that of false positives (FP), false negatives (FN), and true negatives (TN). The first two groups were defined as having, respectively, a non-zero and zero overlap with ground truth (GT). FN predictions were determined as GT segmentations with zero overlap with the predicted lesions. TN examples were obtained by randomly sampling ten sphere-like shapes (with the average lesion volume of the test set), which have zero overlap with the previous groups and are located in patients' brain and skull. The study suggested that maximum and minimum values of XAI maps (with respect to FLAIR) of the four groups present different distributions, as shown in Fig.~\ref{violinplots}.

Radiomic features have been extensively used in radiology to perform a quantitative analysis of medical images \cite{ibrahim2021,annunziata2023,elmahdy2023}. A recent work \cite{ghezzo2023} showed that the outputs of an automatic segmentation model can be used to determine the region of interest (ROI) to extract radiomic features from PET images of intraprostatic cancer lesions.

Inspired by these results, in this work we investigate the discriminatory power of radiomic features extracted from XAI maps to distinguish observations from the first two groups (TP and FP), aiming at improving the model's performance.
Our main hypothesis is that XAI maps contain specific signatures that are distinctive of TP versus FP. This is based on experiments in \cite{spagnolo2024}, which highlight the fact that TPs, FPs, and FNs presented a different distribution of values in XAI maps. The quantitative nature of those saliency maps leads us to test the aforementioned hypothesis. 

\section{Material and methods}

\subsection{Dataset and model}

We used 4023 FLAIR and MPRAGE MRI scans (SIEMENS Avanto/Espree/Symphony 1.5T and Prisma/Skyra/Verio/MAGNETOM Vida 3T, 1mm isotropic) from 687 patients diagnosed with MS (age=45.2±12.2, 433 females, Expanded Disability Status Scale median of 2.5 [0-9]), acquired at the University Hospital of Basel, Switzerland~\cite{SMSC}. WM lesions annotation was performed by three expert clinicians at baseline and follow-ups.
Data were randomly split into training, validation and test sets, containing 560, 90 and 37 patients with 3369, 553 and 101 visits respectively (training/validation set's and test set's mean lesions number of 52.9±36.4 and 42.3±21.4 per patient). Datasets from a same patient were ensured to be included in the same split.

Images were pre-processed by registering FLAIR images to MPRAGE space with the \textit{elastix} toolbox~\cite{klein2009,shamonin2014}, correcting for bias field inhomogeneity~\cite{tustison2010}, and standardizing intensities using z-score.

With the described data, a 3D U-Net~\cite{cicek2016} was trained and tested to segment MS plaques, using patches of dimensions $96^{3}$ and a linear combination of normalized Dice~\cite{raina2023} and blob~\cite{kofler2022} loss. This last choice was made to minimize the impact of instance imbalance within a class and bias towards the occurrence of positive class~\cite{maier-hein2022}.
The trained model achieved a Dice score of 0.60, and a normalized Dice score of 0.71 on the test set, for what concerns lesion segmentation. The model predicted 3050 TP, 1818 FP, and 789 FN examples, reaching an F1 score of 0.7006 and a positive predictive value (PPV) of 0.6265.

\subsection{Instance-level saliency}

Following~\cite{spagnolo2024}, we referred to the lesion domain as $\Omega$: a subset of the image domain $\Gamma$, such that $\Omega\subset\Gamma\subset\mathbb{Z}^D$. Each lesion domain presents a cardinality $|\Omega |$, which is the number of voxels within the lesion. For a given $\Omega$, the generation of explainable maps followed these steps:
\begin{enumerate}
    \item Injecting Gaussian noise $\mathcal{N}(0,\sigma)$ with standard deviation $\sigma=0.05$ to obtain $N$ noisy versions of the input,
    \item Generating a collection of saliency maps for all output voxels in the domain $\Omega$ of the lesion,
    \item Determining the voxel-wise maximum with sign from this collection of maps,
    \item Repeating steps 1-3 and combining these $N=50$ saliency maps to obtain a single one for the target lesion.
\end{enumerate}

The computation of instance-level saliency maps $M_{\Omega}^{\text{gradient}}[\boldsymbol{v}]\in\mathbb{R}$ is summarized in Eq.\eqref{IES}. Originally, separate saliency maps were obtained for each input modality (FLAIR and MPRAGE) to differentiate their respective contribution. For this work we selected maps with gradients computed with respect to FLAIR, based on findings reported in~\cite{spagnolo2024}. The first is that MPRAGE was shown to have a lower contribution to the segmentation of lesions. Secondly, gradients with respect to MPRAGE were less sensitive to the targeted groups (TP, FP, etc.). 

\begin{equation} \label{IES}
    M_{\Omega}^{\text{gradient}}[\boldsymbol{v}] = \frac{1}{N} \sum_{n=1}^{N}D^{n}_{argmax_{\boldsymbol{v}'}|D^{n}_{\boldsymbol{v}'}|}, \text{ where } D^{n}_{\boldsymbol{v}'} = \frac{\partial y(x_{n})[\boldsymbol{v}']}{\partial x_{n}[\boldsymbol{v}]}
\end{equation}

\subsection{Saliency domain-shift verification between training and test sets}

Such explainable maps were generated for TP and FP predictions in the whole test set (3050 and 1818) and a subset \textit{S} of the training set (containing 11569 TP and 9434 FP from 217 and 439 visits). The mean, maximum and minimum values (and their standard deviation) of maps in the training set were compared with those in the test set, to exclude possible domain shifts. 
Such shifts could jeopardize the ability of the radiomics-based model to distinguish between TP and FP on the test set due to changes in the saliency maps.
A Mann-Whitney \textit{U} test was run over the distributions of the three measurements in the two groups.

\subsection{Radiomics feature extraction}

\begin{figure}[!ht]
\includegraphics[width=\textwidth]{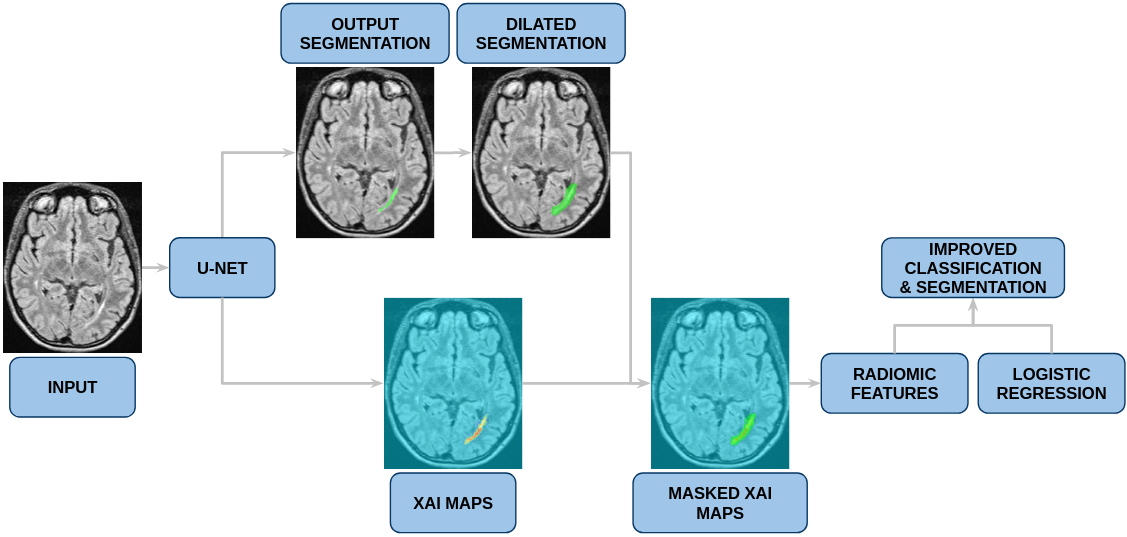}
\caption{Block diagram describing how an XAI map is used to extract radiomic features. In this example, a true positive lesion is shown in the axial plane.}\label{block_diagram}
\end{figure}

Standardized (z-score) explainable maps and the predicted binary lesion masks from \textit{S} were used as input to a radiomic features extractor. The binary segmentation of each lesion was dilated to ensure that the computation of features was determined by the saliency values within the ROI (i.e., the lesion domain $\Omega$ and in its neighborhood). This is needed to avoid the exclusion of negative saliency map values, as described in~\cite{spagnolo2024}. An illustration is provided by the block diagram in Fig.~\ref{block_diagram}.

We used the library \textit{pyradiomics}~\cite{pyradiomics}, with $binWidth=10$ and $sigma=[1,2,3]$. From the pool of 107 available features, those related to \textit{shape} were not considered relevant as depending on the initial segmentation of $\Omega$, resulting in 93 target features.

\subsection{TP/FP prediction refinement}

The radiomic features of saliency maps from the training set were standardized and used to train a logistic regression (LR) model using \textit{scikit-learn}~\cite{scikit-learn}, with the following parameters: maximum number of iterations: 10000, solver: \textit{liblinear}, penalty: \textit{L1}, class weights: 0.29, 0.71. The class weights were selected to reflect the proportion of TP and FP examples per visit in the training set \textit{S}.
To investigate feature importance, we compared the normalized (0-1) regression coefficients of each feature.

The trained model was applied to the entire test set. Following a bootstrapping approach with 1000 iterations, we computed the mean and 95\% confidence interval (CI) of the F1 score and PPV. The updated number of TP, FP and FN predictions was also derived for comparison with the performance of the initial U-Net model. The confusion matrix, F1 score, and PPV reported by the U-Net represent the best scores achieved on the entire test set.

\section{Results}

\subsection{Domain-shift of saliency between training and test sets}
\label{DScheck}

\begin{figure}[!ht]
\includegraphics[width=\textwidth]{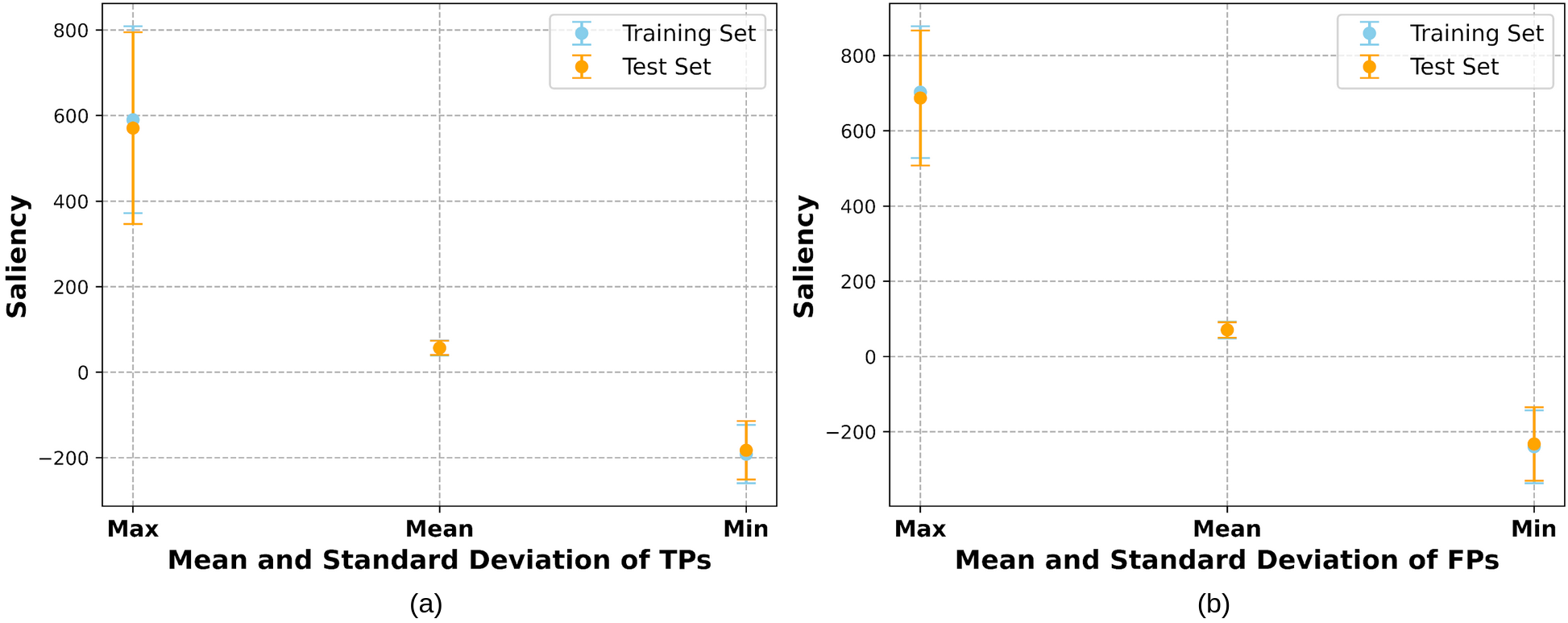}
\caption{Comparison between mean, maximum and minimum values of saliency maps computed on the training and test set, for TP (a) and FP (b) examples.}\label{compare-traintest}
\end{figure}

The comparison between saliency maps computed on the subset \textit{S} of the training set and the test set is summarized in Fig.~\ref{compare-traintest}. For all the metrics, mean values from both sets fell into the interval $\pm \sigma$ (one standard deviation), associated to non-significant differences.

\subsection{Refinement of TP/FP predictions}
\label{refinement}

Testing the LR model with bootstrapping we obtained an F1 score of 0.7450 with a 95\% CI of [0.7358, 0.7547], and a PPV of 0.7817 with a 95\% CI of [0.7679, 0.7962]. The F1 score and PPV of the non-refined U-Net were 0.7006 and 0.6265, respectively.
%, leading to 2732 TP, 763 FP, and 1107 FN predictions. 
A comprehensive performance comparison is presented in Table~\ref{results-table}. 

\begin{table}
\caption{Comparing MS lesion detection performance between the original U-Net and its refined version using logistic regression (LR) relying on radiomic features from XAI maps. We report the number of true positives (TPs), false positives (FPs), false negatives (FNs), and the mean and confidence interval (CI) of F1 score and positive predictive value (PPV).}\label{results-table}
\centering
\begin{tabular}{|c|c|c|}
\hline
& \textbf{U-Net only} & \textbf{U-Net + saliency [95\% CI]} \\
\hline
\textbf{TPs} & 3050 & 2732 \\
\textbf{FPs} & 1818 & 763 \\
\textbf{FNs} & 789 & 1107 \\
\textbf{F1 score} & 0.7006 & 0.7450 [0.7358, 0.7547] \\
\textbf{PPV} & 0.6265 & 0.7817 [0.7679, 0.7962] \\
\hline
\end{tabular}
\end{table}

\begin{figure}[!ht]
\includegraphics[width=\textwidth]{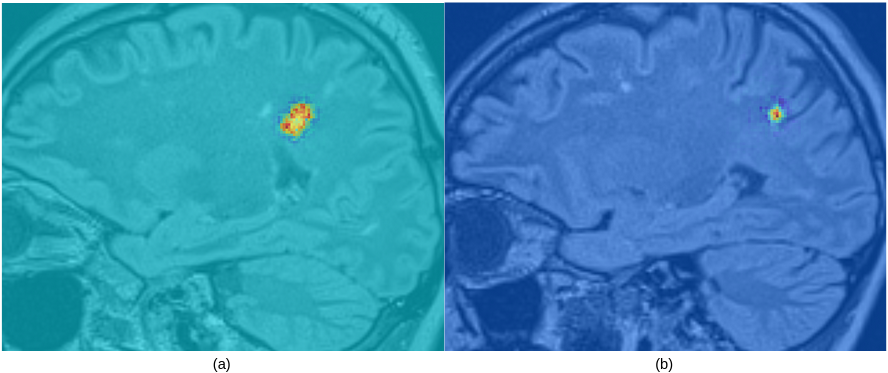}
\caption{An example of a slice in the sagittal plane from a saliency map computed on a true (a) and false (b) positive lesion, scoring 0.9398 and 0.0232 for the true positive class.}\label{example}
\end{figure}

Examples of a slice in the sagittal plane of the 3D XAI maps generated from a TP and a FP are illustrated in Fig.~\ref{example}. The LR output probability (relative to the TP class) for the TP lesion was 0.9398, and 0.0232 for the FP. %The volume and count (median $\pm 95\% CI$) of TP lesions are $42 \pm 22.67$ and $30.5 \pm 3.41$, while in FP predictions are $13 \pm 3.88$ and $15.5 \pm 2.99$.
The example FP candidate is located at the boundary between WM and cortex, and corresponds to one of the brain sulci.

\begin{figure}[!ht]
\includegraphics[width=\textwidth]{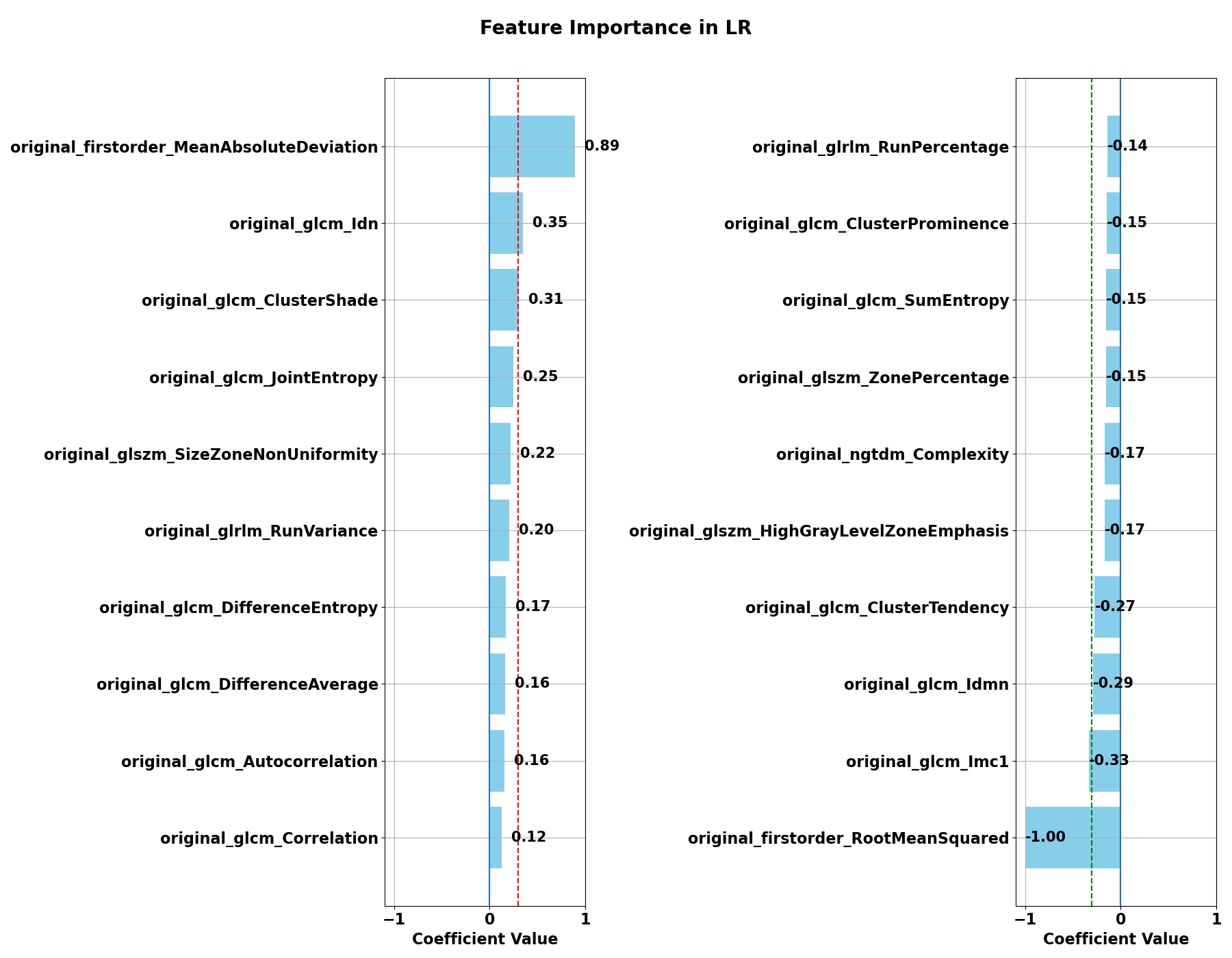}
\caption{Normalized radiomic features showing the highest importance (top 10 positive on the left, top 10 negative on the right), in terms of LR coefficients. The dashed red line represents a coefficient value of 0.3, the dashed green line a coefficient value of -0.3.}\label{feature_importance}
\end{figure}

The importance of radiomic features used by the model is presented in Fig.~\ref{feature_importance}. The most important were two features based on saliency intensity, mean absolute deviation (MAD) and Root Mean Squared (RMS), and reported a normalized coefficient of 0.89 and -1.0 respectively.  

\section{Discussion}

% results, trade off using only max/min and after
This work employed radiomics to extract features from XAI (saliency) maps generated for a DL semantic segmentation model. These features were fed to a linear model (i.e., LR) to discriminate between FP and TP predictions, and improve the classification score.
Our main hypothesis was that XAI maps contained specific signatures that are distinctive of TP versus FP.
As a first step of our investigation, we tried using only the minimum and maximum values of saliency to discriminate between TP and FP classes. In this way, the binary classifier could reach a higher precision but a lower recall, eventually resulting in a trade off where the F1 score did not improve. 
A similar trend was observed when trying to adjust the segmentation threshold at the U-Net output layer. 
The results in Table~\ref{results-table} show that using radiomics on saliency maps, it was possible to refine classification scores, improving the F1 score from 0.70 to 0.75 and the PPV from 0.63 to 0.78. Confidence intervals computed for the refined model are not including the performance values of the initial model, suggesting significance of the observed improvements.

% no domain shift between train and test images
In Section \ref{DScheck} we compared the maximum, minimum and mean values of XAI in the training and test set. According to this analysis, we found no evidence of domain shift between the two sets. The characteristic of having a slightly positive mean value was also preserved in both groups (TP and FP) and sets. This indicated that the radiomics model trained on saliency maps obtained on the training set will generalize to those of the test set.

% Interpretation of relevant features
In Section \ref{refinement} we reported the features which contribute the most to discriminate between saliency maps from TP and FP examples. A strong positive coefficient for MAD reveals that the intensity variability with respect to the mean is higher for XAI maps of the TP class. This finding supports prior results obtained on maximum, minimum and mean values. A positive Inverse difference normalized (Idn) may indicate a more homogeneous texture surrounding voxels. A strong negative coefficient for RMS may describe saliency maps for the FP class as highly variable, and presenting more outliers. A negative Informational measure of correlation (Imc1) represents a higher tendency of neighbouring voxel pairs in FP cases to have similar values. Such findings suggest that saliency maps for TP examples may present a wider range of values (contrast between positive and negative regions), less outliers, and overall a more homogeneous texture.  

% limitations, including FNs for next?
This work has also some limitations. 
The application of this method is limited to identifying and ruling out FPs. A possible future improvement would be to extend the approach to FNs, and to refine the detection performance even further. In that case the issue would be to have FN candidates on the test set without using a ground truth. For this purpose, an estimation of the uncertainty~\cite{nair2020,molchanova2023} of the model's prediction could help to target possibly missed lesions.
Moreover, this study explores the refinement of predictions from a single DL model. The impact of XAI on different models should be investigated, as we would expect the initial performance of the model to be relevant.
In addition, it would be important to evaluate the performances of this method on saliency maps coming from an out of domain test set, or training and testing the model on multiple datasets. Though, this would also require XAI maps to be comparable for multi-centric data which, to our knowledge, has not been investigated yet. Nevertheless, it is important to remark that the MRI data used in this study were acquired with multiple scanners.
Furthermore, investigating the which and how many radiomic features are required to achieve a desired performance would be of interest.
 
\section{Conclusion}

This work demonstrated that radiomic features extracted from XAI (saliency) maps generated for a DL semantic segmentation model can be used to improve the detection performance of a segmentation model by a large margin.

\begin{credits}
\subsubsection{\ackname} This work was supported by the Hasler Foundation with the project MSxplain number 21042, the Swiss National Science Foundation (SNSF) with the project 205320\_219430, and the Swiss Cancer Research foundation with the project TARGET (KFS-5549-02-2022-R). We acknowledge access to the expertise of the CIBM Center for Biomedical Imaging, a Swiss research center of excellence founded and supported by CHUV, UNIL, EPFL, UNIGE and HUG.

\subsubsection{\discintname}
The University Hospital Basel (USB) and the Research Center for Clinical neuroimmunology and Neuroscience (RC2NB), as the employers of Cristina Granziera, have received the following fees which were used exclusively for research support from Siemens, GeNeuro, Genzyme-Sanofi, Biogen, Roche. They also have received  advisory board and consultancy fees from Actelion, Genzyme-Sanofi, Novartis, GeNeuro, Merck, Biogen and Roche; as well as speaker fees from Genzyme-Sanofi, Novartis, GeNeuro, Merck, biogen and Roche. Federico Spagnolo was an employee of F. Hoffmann-La Roche Ltd. The remaining authors have no competing interests to declare that are relevant to the content of this article.
\end{credits}


\end{document}